# Isotropic superconducting gaps with enhanced pairing on electron Fermi surfaces in FeTe$_{0.55}$Se$_{0.45}$

H. Miao,[1] P. Richard,[1] Y. Tanaka,[2] K. Nakayama,[2] T. Qian,[1] K. Umezawa,[2] T. Sato,[2,3] Y.-M. Xu,[4] Y.-B. Shi,[1] N. Xu,[1] X.-P. Wang,[1] P. Zhang,[1] H.-B. Yang,[5] Z.-J. Xu,[5] J. S. Wen,[5] G.-D. Gu,[5] X. Dai,[1] J.-P. Hu,[1,6] T. Takahashi,[2,7] H. Ding[1]

[1]Beijing National Laboratory for Condensed Matter Physics, and Institute of Physics, Chinese Academy of Sciences, Beijing 100190, China

[2]Department of Physics, Tohoku University, Sendai 980-8578, Japan

[3]TRiP, Japan Science and Technology Agency (JST), Kawaguchi 332-0012, Japan

[4]Materials Sciences Division, Lawrence Berkeley National Laboratory, Berkeley, California 94720, USA

[5]Condensed Matter Physics and Materials Science Department, Brookhaven National Laboratory, Upton, New York 11973, USA

[6]Department of Physics, Purdue University, West Lafayette, Indiana 47907, USA

[7]WPI Research Center, Advanced Institute for Materials Research, Tohoku University, Sendai 980-8577, Japan

**Abstract:**

The momentum distribution of the energy gap opening at the Fermi level of superconductors is a direct fingerprint of the pairing mechanism. While the phase diagram of the iron-based superconductors promotes antiferromagnetic fluctuations as a natural candidate for electron pairing, the precise origin of the interaction is highly debated. We used angle-resolved photoemission spectroscopy to reveal directly the momentum distribution of the superconducting gap in $FeTe_{1-x}Se_x$, which has the simplest structure of all iron-based superconductors. We found isotropic superconducting gaps on all Fermi surfaces whose sizes can be fitted by a single gap function derived from a strong coupling approach, strongly suggesting local antiferromagnetic exchange interactions as the pairing origin.

Superconductivity occurs at a critical temperature $T_c$ below which electron pair formation induces an energy gap whose size and momentum ($k$) dependence is determined by the pairing mechanism. While lattice vibrations are known to mediate superconductivity in ordinary metals, electronic spin fluctuations are widely believed to be critical to superconductivity in the new iron-based high-temperature superconductors due to the proximity between the superconducting (SC) phase and the magnetic ordered phase. The driving mechanisms of both magnetic ordering and superconductivity is currently under intensive debate, mainly between two approaches: a weak coupling approach where both magnetic and SC orders are driven by the enhanced spin susceptibility near the Fermi surface (FS), and a strong coupling approach where the two orders are caused by local magnetic exchange interactions between neighboring electrons. The weak coupling approach has found support from the observation of quasi-nesting between hole FS pockets at the Brillouin zone (BZ) center and electron FS pockets located at the antiferromagnetic wave vector in most ferropnictides (*1-5*). However, this approach has encountered difficulty in explaining the bi-collinear magnetic ordering pattern in the ferrochalcogenide FeTe (*6*). Moreover, the weak coupling approach was seriously challenged by the recent observation of isotropic SC gaps in a new ferrochalcogenide superconductor, $A_xFe_2Se_2$ (A = K, Rb, Cs, Tl), without FS nesting on the particle-hole channel (*7-9*). On the other hand, the strong coupling approach (e.g., $J_1$-$J_2$-$J_3$ model) fares better in explaining the two examples above (*10, 11*).

Precise information on the pairing strength along different FS pockets of ferrochalcogenide superconductor $FeTe_{1-x}Se_x$ can be used to further distinguish these two approaches, due to its close relationship to its magnetic parent FeTe and SC cousin $K_xFe_2Se_2$. While a previous angle-resolved photoemission spectroscopy (ARPES) on $FeTe_{0.7}Se_{0.3}$ indicates an isotropic gap on the holelike FS pocket (*12*), angle-resolved specific heat (ARSH) experiment under high magnetic field claimed nodes or minimal gaps in a $FeTe_{0.55}Se_{0.45}$ sample (*13*). Nodes were also suggested from STM measurements on FeSe crystalline films (*14*), whereas a STM study on $FeTe_{1-x}Se_x$ single-crystals ($T_c$ = 13 to ~14.5 K) favors a nodeless $s_\pm$ pairing symmetry (*15*). Here we report high-resolution ARPES results on high-quality $FeTe_{0.55}Se_{0.45}$ SC samples ($T_c$ = 14.5 K) [see supporting online material]. We observe isotropic SC gaps on both electronlike and holelike FS pockets. More importantly, the pairing strength is found to be stronger on the electronlike FSs, which can be naturally explained by the enhanced antiferromagnetic exchange between the 3$^{rd}$ nearest neighbors ($J_3$) in this material. All the different SC gaps can be fitted by a single gap function that is fully consistent with the strong coupling $J_1$-$J_2$-$J_3$ approach.

We start with the ARPES measurement of a cut along the Γ-M (π, 0) direction, here defined in the 1 Fe per unit cell notation. Figs. 1A and 1C show the ARPES intensity plots along this cut in the normal state (25 K) and SC state (6 K), respectively. The corresponding energy distribution curves (EDCs) are given in Figs. 1B and 1D, respectively. The normal state data show clearly two intense holelike bands at the Γ point. While the inner one (α) sinks ~ 14 meV below $E_F$, the outer one (α') crosses $E_F$ very close to Γ. A third weaker holelike band with a larger Fermi wavevector ($k_F$) and a larger effective mass, named β band, is also visible. Upon cooling temperature below $T_c$

= 14.5 K, coherent peaks associated with superconductivity develop on the α' and β bands, as well evidenced by the low-temperature EDCs shown in Fig. 1D. The temperature evolution of the coherent peak is drastic and confirms their SC origin. The temperature dependence of the EDC recorded at a particular $k_F$ of the β band is displayed in Fig. 1E, which shows dramatic lineshape change across $T_c$. The impressive sharpness of these SC coherent peaks (full-width-at-half-maximum < 5 meV) compared to a previous ARPES study (*12*) on $Fe_{1.03}Te_{0.7}Se_{0.3}$ indicates the improved quality of the single-crystals used in the current study.

An additional SC gap is found on the electronlike band (γ) observed at the M point. Fig. 1F compares EDCs recorded in the SC state at the $k_F$ of the α', β and γ bands. The SC gap size on these three bands is different. We observe gap sizes of 1.7, 2.5 and 4.2 meV for the α', β and γ bands, respectively. Although multigap superconductivity has already been reported in other iron-based systems (*1-4, 12, 16-17*), it is the first time that the SC gap at the M point is the largest SC gap observed, with the exception of the 122-chalcogenides, which do not have Γ-centered holelike FSs at all (*7-9, 18*).

Whether gap nodes exist at the M point or not in the SC state is a highly debated issue in the study of iron-based superconductors. A previous ARSH experiment claims nodes or minimal gaps in $FeTe_{0.55}Se_{0.45}$ samples (*13*). To check this point we performed high-energy resolution *k*-dependent measurements on both Γ- and M-centered FS pockets. Figs. 2A and 2B show the symmetrized EDCs of the β and γ bands, respectively, at different $k_F$ positions indicated on the FS displayed in Fig. 2C. As illustrated by the polar plot in Fig. 2D, the gap sizes on both pockets are nodeless and quite isotropic, with error bars limiting possible anisotropy to 15%.

It has been proposed that the orbital character may play an important role in determining the pairing strength and the symmetry for this multi-orbital system (*19*, *20*). To address this issue, we have performed polarization-dependent ARPES measurements along high symmetry lines to gain information on the orbital symmetry. Figs. 3A and 3B show intensity plots along Γ–M under *S*-polarization and *P*-polarization, respectively. We observed that the α' band is enhanced under *S*-polarization, but is significantly suppressed by *P*-polarization. The α band, however, has reversed response under *S* and *P* polarizations. The same behavior has been observed in a previous ARPES measurement for α and α' bands along Γ–X (*21*). These behaviors suggest that the α (α') band has an even (odd) symmetry along both Γ–M and Γ–X. Similarly, we determine that the β band is even along Γ–X but odd along Γ–M, in agreement with previous reports (*21*, *22*). Taking LDA calculations as reference (*21-23*), we reach the following conclusion on the orbital characters for the three bands observed around Γ, which is summarized in Fig. 3C: the β band has mainly a $d_{xy}$ orbital character whereas the α (α') band is mainly the even (odd) combination of the $d_{xz}$ and $d_{yz}$ orbitals (labeled as $d_{even}$ ($d_{odd}$)). Therefore, we conclude that the 2.5 and 1.7 meV gaps at Γ originate from the $d_{xy}$ and $d_{xz}/d_{yz}$ orbitals, respectively. As mentioned above, the α band top is 15 meV away from $E_F$ and is thus slightly separated from the α' band top. This contradicts the LDA expectation of degeneracy of the $d_{even}$ and $d_{odd}$ bands at Γ preserved by the 4-fold symmetry. Careful examination from LDA calculations reveals a possible Se/Te $p_z$ band situation around the

BZ center, complicating the band dispersion around Γ. We note that the small gap of 1.7 meV at the Γ point may be related at least partly to the Se/Te $p_z$ orbital, rather than purely to the Fe3$d$ orbitals that generate stronger pairing.

Our ARPES results on FeTe$_{0.55}$Se$_{0.45}$ are in good agreement with other experimental measurements on this material. Fig. 3D shows comparison of our ARPES EDC spectra and the STM conductance curve measured on similar FeTe$_{1-x}$Se$_x$ ($T_c$ ~ 14.5 K) (*15*). It is clear that all the three gaps observed by ARPES can find corresponding features in the STM data. The smallest gap of 1.7 meV observed at Γ matches well the 1.8 meV peak in the STM curve. The 4.2 meV gap observed on the electron FS and the 2.5 meV gap on the β hole FS correspond well to the 4 meV peak and the weak 2.5 meV shoulder in the STM curve, respectively.

We further compare our results with optical measurements (*24*) on the same sample batch, as shown in Fig. 3E. The later is a well-known bulk sensitive technique. We find a good consistency of gap size that excludes surface contribution in our observations. It is important to point out that samples cleave between two equivalent weakly-bonded Se(Te) planes, exposing a non-polar surface which is usually bulk-representative. The bulk nature is also reflecting from the ARPES observation that the SC gap and coherence diminish right at the bulk $T_c$, as demonstrated by Fig. 1E. In addition, inelastic neutron scattering has revealed a resonant mode around 6 - 7 meV in the SC state of samples with similar $T_c$ and composition (*25*). We note that the sum of the gaps on the β and γ FSs (6.7 meV) is compatible to the energy of the resonant mode, which indicates that this resonant mode is related to scattering between the β hole FS and the γ electron FS.

Assuming that high magnetic field does not change the topology of the FS and the SC gap function, the contrast between highly anisotropic gap suggested by ARSH experiments (*13*) and isotropic gaps observed by ARPES is puzzling. Even though nodes at a particular $k_z$ is still possible, the quasi-two dimensionality of this FeTe$_{1-x}$Se$_x$ material makes this scenario unlikely. Another possible cause is that the cleaved surface may have enhanced impurity scattering tending to average out possible gap anisotropy, which is also unlikely since isotropic gaps have been observed by ARPES for different structures of iron-based superconductors with different surfaces exposed. We note that all the SC coherent peaks in this material are very sharp, although the ones on the electronlike pocket are not as pronounced as the ones on the holelike pocket, indicating weak scattering effect. It is worth to point out that small linewidth broadening induced by impurities does not usually shift the quasiparticle peak position, or the gap value observed by ARPES. However, it may affect some thermodynamics measurements that are sensitive to the residual density of states. An alternative explanation to account for the discrepancy between conclusions reached from ARPES and ARSH experiments is that transport measurements may be more sensitive to the overall gap distribution, which is anisotropic with respect to the Γ point because of the FS topology itself, rather than to the detail of the gap distribution around each FS taken separately.

We discuss a peculiar gap size "anomaly" observed in this material: the gap amplitude is stronger on the electron FS than on the hole FS, as shown in Fig. 4A. This is quite different from ferropnictide superconductors, which have similar gap size on the "quasi-nested" holelike and electronlike FSs (*1-3, 17*). Similar gap values on similar-size FSs can be captured by the simple $s_\pm$ gap function $\cos(k_x)\cos(k_y)$. Such gap function is naturally derived from the effective $J_1$-$J_2$ model (*26*), where $J_1$ and $J_2$ are the nearest- and next-nearest-neighbor magnetic exchange interaction strengths, respectively. It is clear from Fig. 4B that a simple $\cos(k_x)\cos(k_y)$ function cannot fit the gap on both hole and electron FSs. However, we notice that the exchange parameters for ferrochalcogenides are different from the ones of ferropnictides, in such a way that (i) $J_1 < 0$ (ferromagnetic) and (ii) $J_3$ (next-next-nearest-neighbor exchange) is no longer negligible. A sizable AF $J_3$ (~ 7 meV) was reported from an inelastic neutron scattering measurement on FeTe (*10*), which is believed to play a critical role in forming the bi-collinear magnetic pattern in FeTe. It is known that the *s*-wave pairing form induced by AF $J_3$ is $\Delta_3(\cos(2k_x)+\cos(2k_y))/2$. Combined with $\cos(k_x)\cos(k_y)$ induced by AF $J_2$, the pairing form in this material should be $|\Delta_2\cos(k_x)\cos(k_y)-\Delta_3(\cos(2k_x)+\cos(2k_y))/2|$. Indeed, as demonstrated in Fig. 4C, this gap function fits all the gaps reasonably well, with $\Delta_2 = 3.55$ meV and $\Delta_3 = 0.95$ meV. The ratio between $\Delta_2$ and $\Delta_3$ is similar with $J_2/J_3$ (22/7) given by a neutron study (*10*). We note that the small gap of 1.7 meV at $\Gamma$ would fall slightly off the fitting curve, as expected if the $\alpha$' band hybridizes with Se$4p_z$ states.

Stronger and isotropic pairing on the electron FS sheets of FeTe$_{1-x}$Se$_x$ has important implications to superconductivity in another ferrochalcogenide, AFe$_2$Se$_2$ (A = K, Rb, Cs, Tl), with $T_c$ as high as 31 K (*27, 28*). It has been also observed by several ARPES studies (*7-9*) that the SC gap on its electron FS is also isotropic. Interestingly, the

$2\Delta/T_c$ ratios are similar in these two ferrochalcogenides [7 in AFe$_2$Se$_2$ (*7*) *vs* 6.7 in FeTe$_{1-x}$Se$_x$]. A recent neutron study revealed a large $J_3$ (~ 9 meV) for K$_{0.8}$Fe$_{1.6}$Se$_2$ as well (*29*). Thus we propose that AFe$_2$Se$_2$ and FeTe$_{1-x}$Se$_x$ belong to a same class as far as superconductivity is concerned, with the same pairing function: $|\Delta_2\cos(k_x)\cos(k_y)-\Delta_3(\cos(2k_x)+\cos(2k_y))/2|$. As shown in Fig. 4D, the enhanced pairing on the electron FS around M accompanies the reduced pairing on the hole FS around $\Gamma$. For AFe$_2$Se$_2$, which does not have hole FS at $\Gamma$, this reduction of pairing may have no effect on $T_c$. Clearly, the observation of isotropic SC gaps with strong pairing on the electronlike FS pocket in the ferrochacogenides supports the strong coupling local pairing picture for iron-based superconductors.

## Materials and methods:

The high-quality single crystals were grown by a unidirectional solidification method with a nominal composition of $FeTe_{0.55}Se_{0.45}$ and a critical temperature determined by magnetic susceptibility of $T_c$ = 14.5 K with a transition width of 1 K. ARPES measurements were performed at Tohoku University using a VG-Scienta SES 2002 multi-channel analyzer and the He Iα resonance line of a helium discharge lamp (hv = 21.218 eV). Additional data were recorded at the Synchrotron Radiation Center (Stoughton, WI) with a VG-Scienta SES 200 using 52 eV light with tunable polarization, as well as at the National Light Source of Brookhaven National Laboratory using a VG-Scienta SES 2002 multi-channel analyzer with 16 eV photons. The energy resolution was set at 2 meV for the measurement of the SC gap. All preparation procedures for the gap measurements were done in a glove box filled with He gas to avoid contact with air. Clean surfaces were obtained by cleaving samples in situ cleaving in a working vacuum better than $1 \times 10^{-10}$ Torr.

## Acknowledgment:

We acknowledge B. A. Bernevig, A. Chubukov, Z. Fang, D.-H. Lee, H.-H. Wen and G. Xu for valuable discussions, and T. Kawahara for his help in the ARPES experiments. We also thank T. Hanaguri and C. Homes for providing STM and optical conductivity data, respectively, for comparison with our ARPES results. This work was supported by grants from Chinese Academy of Sciences (2010Y1JB6), Ministry of Science and

Technology of China (2010CB923000), Chinese National Science Foundation (11004232 and 11050110422), JSPS, TRiP-JST, MEXT of Japan, CREST-JST and National Science Foundation (Contract No. DE-AC02-05CH11231). This study is based in part upon research conducted at the Synchrotron Radiation Center, University of Wisconsin-Madison, and at the National Light Source of Brookhaven National Laboratory, which are supported by the National Science Foundation under Award No. DMR-0537588, and by the Department of Energy of USA under contract No. DE-AC02-98CH10886, respectively.

**Figure legends**

**Fig. 1.** (**A**) ARPES intensity plot along Γ-M cut in the normal state ($T = 25$ K). The blue and black dotted lines indicate the $k_F$ positions of the α' and β bands, respectively. (**B**) EDCs of (A). Blue, green and black thick bars mark the peak positions of the α, α' and β bands, respectively. The blue and black EDCs correspond the $k_F$ positions indicated in 1(A). (**C**) and (**D**) Same as (**A**) and (**B**) but in the SC state ($T = 6$ K). (**E**) Temperature dependence of the SC gap of the β band. EDCs show a strong coherent peak below $T_c$ (14.5 K) that is completely suppressed above $T_c$. (**F**) Comparison of the SC gaps of the α', β and γ bands. Black thick bars indicate the peak positions.

**Fig. 2.** (**A**) and (**B**) SC gap momentum dependence of the β and γ bands, respectively. Symmetrized EDCs are used to approximately remove the contribution of the Fermi-Dirac function. The separation between two peaks is equal to 2Δ. (**C**) FS indicating positions where the gap size has been measured. Dotted FSs are folded from the BZ of a pure Fe plane due to the alternate position of the chalcogen atoms above and below the Fe plane. (**D**) Polar plot of the SC gaps with circles indicating the average values on each FS pocket.

**Fig. 3.** (**A**) and (**B**) ARPES intensity plots (20 K) along the Γ-M high-symmetry line measured with $S$ and $P$ polarizations ($h\nu = 52$ eV), respectively. The different intensity

under different polarizations indicates that the α and α' bands have an even and an odd symmetry along Γ-M, respectively. (**C**) Schematic band structure and dominant orbital character of each band at Γ. (**D**) SC gap correspondence between ARPES and STM measurements. Blue, red and green lines are $k_F$ symmetrized EDCs of the α, β and γ bands, respectively. The black curve is the data recorded by STM (*15*). Dashed arrows indicate the correspondence between ARPES and STM features. (**E**) Comparison of ARPES data with optical conductivity (*24*). Colored arrows are positioned at the 2Δ values recorded by ARPES. Dashed lines indicate the 2Δ values determined by ARPES.

**Fig. 4.** (**A**) 3D representation of the SC gap with the FS topology. (**B**) Fit of the SC gap data with a $|\Delta_2\cos(k_x)\cos(k_y)|$ gap function. (**C**) Fit of the SC gap data with a $|\Delta_2\cos(k_x)\cos(k_y)-\Delta_3(\cos(2k_x)+\cos(2k_y))/2|$ gap function. (**D**) In-plane momentum distribution of the $|\Delta_2\cos(k_x)\cos(k_y)-\Delta_3(\cos(2k_x)+\cos(2k_y))/2|$ SC gap function along with the FSs. Color scale shows the absolute gap size.

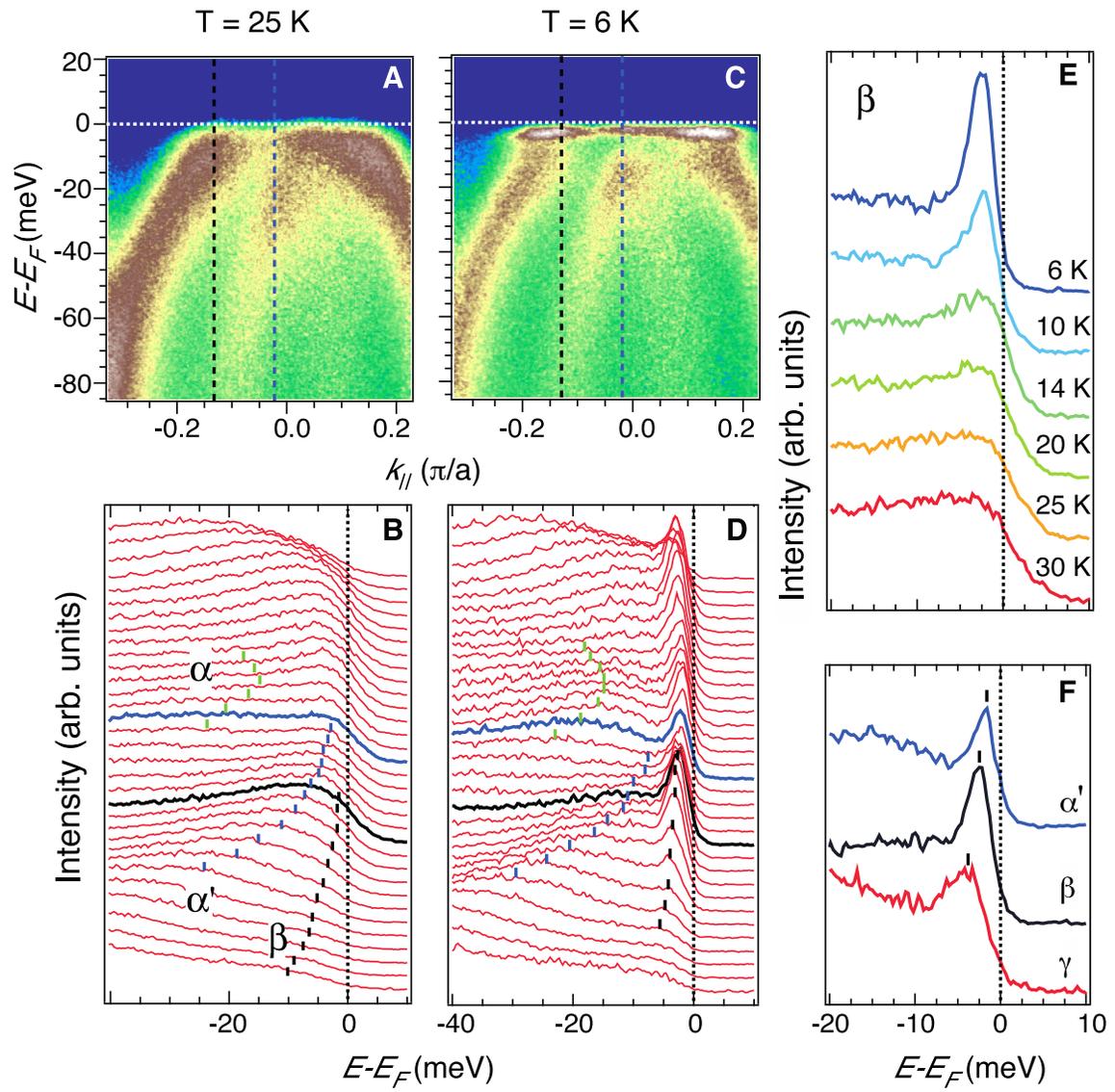

**Fig. 1**

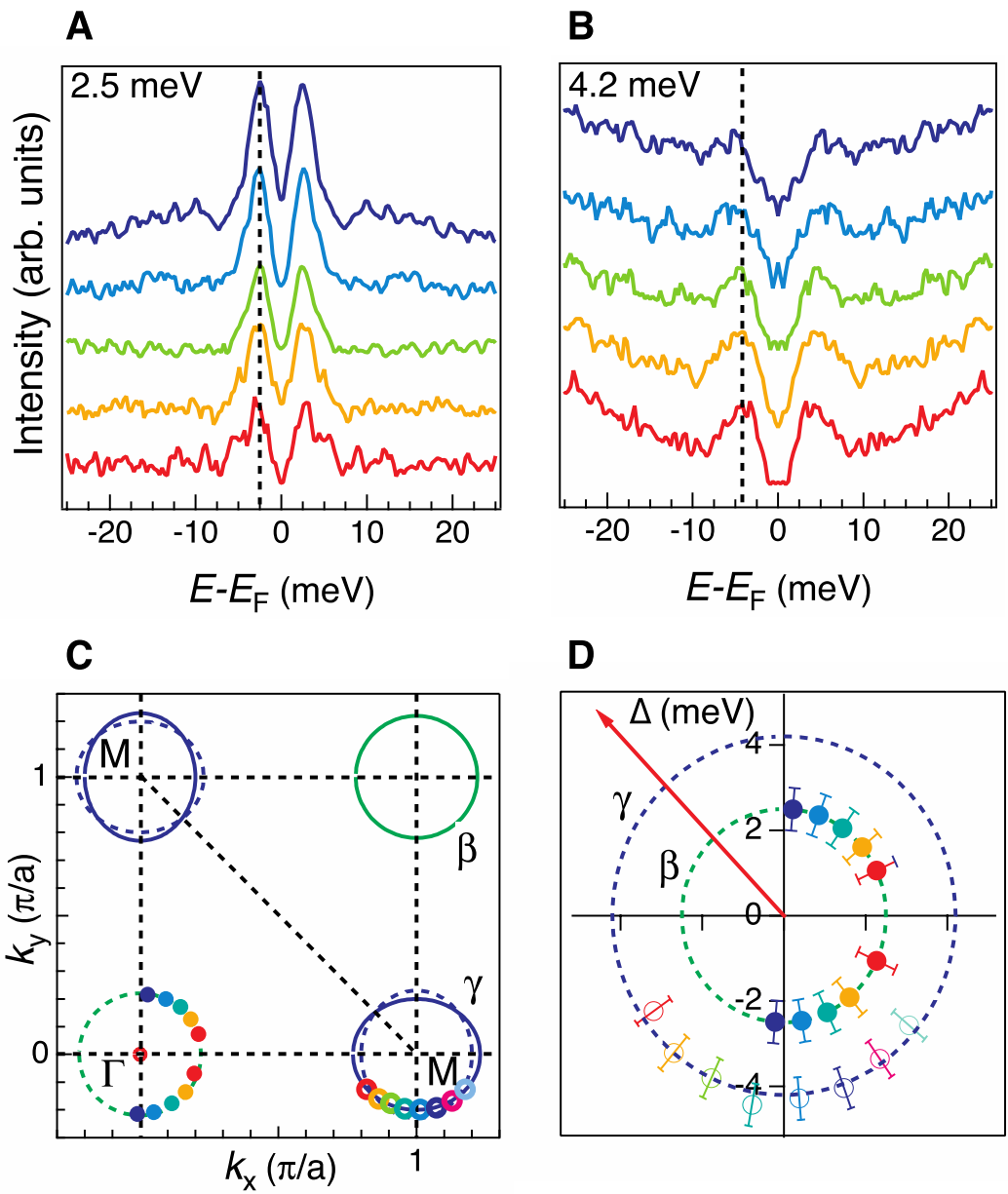

Fig. 2

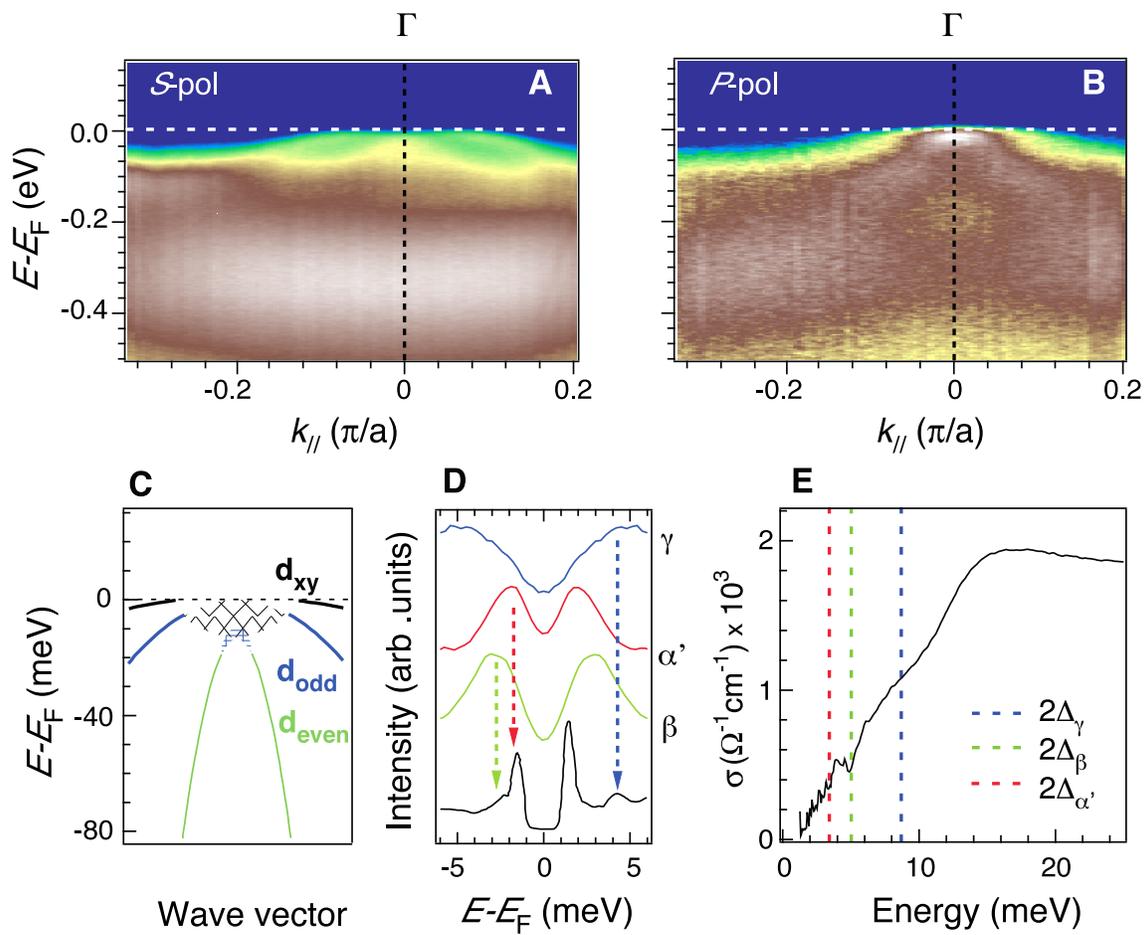

Fig. 3

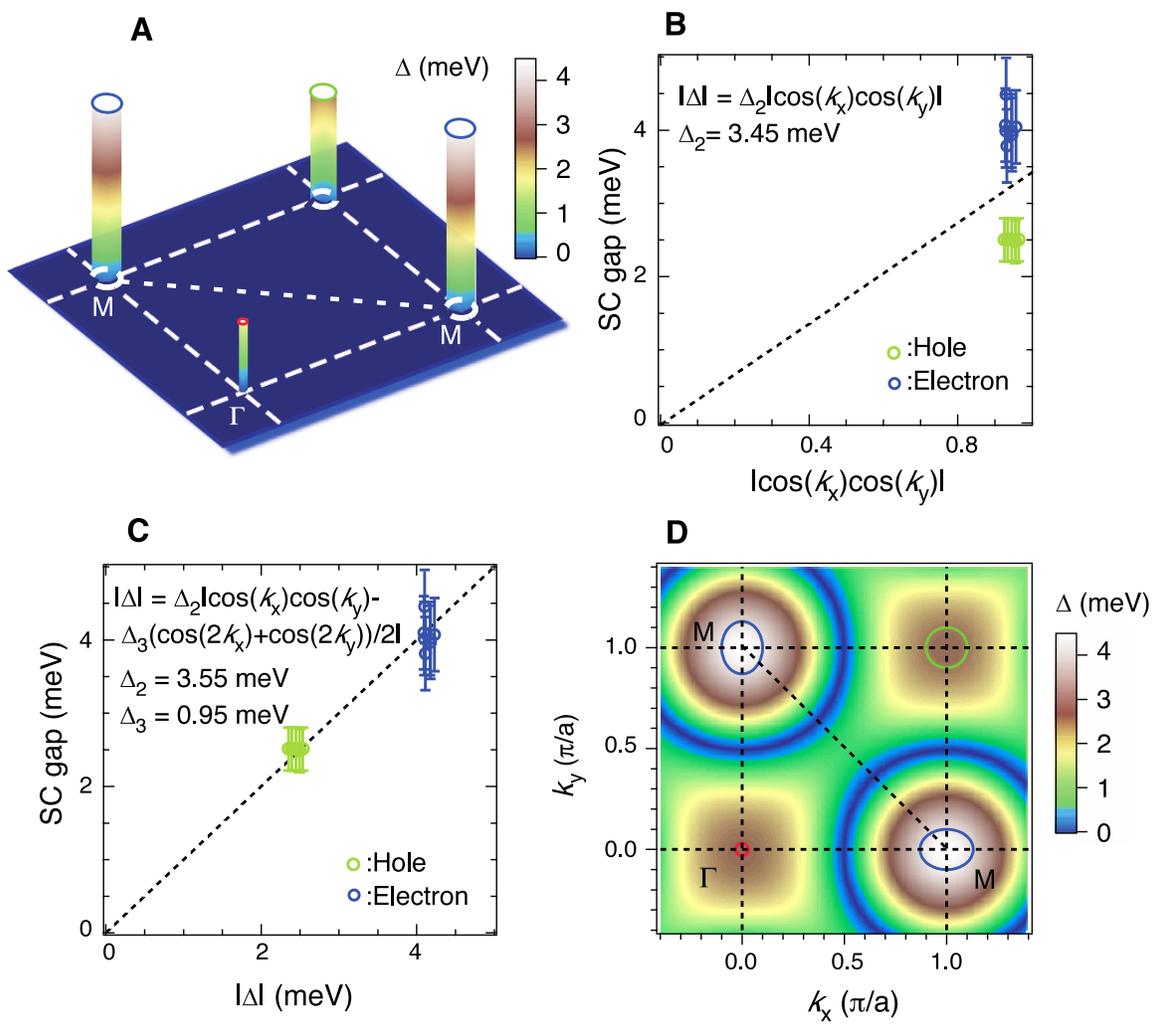

Fig. 4